# To BLOB or Not To BLOB:
# Large Object Storage in a Database or a Filesystem?

Russell Sears[2], Catharine van Ingen[1], Jim Gray[1]
1: Microsoft Research, 2: University of California at Berkeley
sears@cs.berkeley.edu, vanIngen@microsoft.com, gray@microsoft.com

# To BLOB or Not To BLOB:
# Large Object Storage in a Database or a Filesystem?

Russell Sears[2], Catharine van Ingen[1], Jim Gray[1]
1: Microsoft Research, 2: University of California at Berkeley
sears@cs.berkeley.edu, vanIngen@microsoft.com, gray@microsoft.com

## *Abstract*

Application designers must decide whether to store large objects (BLOBs) in a filesystem or in a database. Generally, this decision is based on factors such as application simplicity or manageability. Often, system performance affects these factors.

Folklore tells us that databases efficiently handle large numbers of small objects, while filesystems are more efficient for large objects. Where is the break-even point? When is accessing a BLOB stored as a file cheaper than accessing a BLOB stored as a database record?

The simple answer is: BLOBs smaller than 256KB are more efficiently handled by a database, while a filesystem is more efficient for those greater than 1MB. Of course, this will vary between different databases and filesystems.

By measuring the performance of a storage server that mimics common workloads we found that the break-even point depends on many factors. However, our experiments suggest that *storage age*, the ratio of bytes in deleted objects to bytes in live objects, is dominant. As storage age increases, fragmentation tends to increase. The filesystem we study has better fragmentation control than the database we used, suggesting the database system would benefit from incorporating ideas from filesystem design. Conversely, filesystem performance may be improved by using database techniques to handle many small files.

Surprisingly, for these studies, when average object size is held constant, the distribution of object sizes did not significantly affect performance. We also found that, in addition to low percentage free space, a low ratio of free space to average object size leads to fragmentation and performance degradation.

# 1. Introduction

Application data objects are getting larger as digital media becomes ubiquitous. Application designers have the choice of storing large objects as files in the file system, as BLOBs (binary large objects) in a database, or as a combination of both. Only folklore is available regarding the tradeoffs – often the design decision is based on which technology the designer knows best. Most designers will tell you that a database is probably best for small binary objects and that that files are best for large objects. But, what is the break-even point? What are the tradeoffs?

This article characterizes the performance of an abstracted typical web application that deals with relatively large objects. Two versions of the system are compared; one uses a relational database to store large objects, while the other version stores the objects as files in the filesystem. We measure how the performance changes over time as the storage becomes fragmented. The article concludes by describing and quantifying the factors that a designer should consider when picking a storage system. It also suggests filesystem and database improvements for large object support.

One surprising (to us at least) conclusion of our work is that storage fragmentation is the main determinant of the break-even point in the tradeoff. Therefore, much of our work and much of this article focuses on storage fragmentation issues. In essence, file systems seem to have better fragmentation handling than databases and this drives the break-even point down from about 1MB to about 256KB.

# 2. Background

## *2.1. Fragmentation*

Filesystems have long used sophisticated allocation strategies to avoid fragmentation while laying out objects on disk. For example, OS/360's filesystem was extent based and clustered extents to improve access time. The VMS filesystem included similar optimizations and provided a file attribute that allowed users to request a (best effort) contiguous layout [McCoy, Goldstein]. Berkeley FFS [McKusick] was an early UNIX filesystem that took sequential access, seek performance and other hardware characteristics into account when laying data out on disk. Subsequent filesystems were built with similar goals in mind.

The file system used in these experiments, NTFS, uses a 'banded' allocation strategy for metadata, but not for file contents [NTFS]. NTFS allocates space for file stream data from a *run-based lookup cache*. Runs of contiguous free clusters are ordered in decreasing size and volume offset. NTFS attempts to satisfy a new space allocation from the outer band. If that fails, large extents within the free space cache are used. If that fails, the file is fragmented. Additionally, when a file is deleted, the allocated space cannot be reused immediately; the NTFS transactional log entry must be committed before the freed space can be reallocated. The net behavior is that file stream data tends to be allocated contiguously within a file.

In contrast, database systems began to support large objects more recently [DeWitt]. They historically focused on small (100 byte) records and on clustering tuples within the same table. Clustered indexes let users control the order in which tuples are stored, allowing common queries to be serviced with a sequential scan over the data.

Filesystems and databases take different approaches to modifying an existing object. Filesystems are optimized for appending or truncating a file. In-place file updates are efficient, but when data are inserted or deleted in the middle of a file, all contents after the modification must be completely re-written. Some databases completely rewrite modified blobs; this rewrite is transparent to the application. Others, such as SQL Server, support efficient insertion or deletion within an object if the change is aligned on database pages. (How likely that may be depends on how well the database page aligns to the application change to the object.) Still others, such as Exodus [DeWitt], use B-Tree based storage of large objects, and allow efficient insertion and deletion of arbitrary size data at arbitrary offset within an object.

Of course the application can do its own fragmentation. Some applications simply store each object directly in the database as a single blob, or as a single file in a file system. However, in order to efficiently service user requests, other applications use more complex allocation strategies. For instance, objects may be partitioned into many smaller chunks stored as blobs. Video streams are often "chunked" in this way. Alternately, smaller objects may be aggregated into a single blob or file – for example TAR, ZIP, or CAB files.

### *2.2. Safe writes*

This study only considers applications that overwrite entire objects. Such applications do not force the storage system to understand their data's structure, trading opportunities for optimization for simplicity and robustness. Even this simple update policy is not entirely straightforward. This section will describe mechanisms that safely update entire objects at once.

Most filesystems protect internal metadata structures (such as directories and filenames) from corruption due to dirty shutdowns (such as system crashes and power outages). However, there is no such guarantee for file contents. In particular, file systems and the operating system below them reorder write requests to improve performance. Only some of those requests may complete at dirty shutdown.

As a result, many desktop applications use a technique called a "safe write" to achieve the property that, after a dirty shutdown, the file contents are either new or old but not a mix of old and new. While safe-writes ensure that an old file is robustly replaced, they also force a complete copy of the file to disk even if most of the file contents are unchanged.

To perform a safe write, a new version of the file is created with a temporary name. Next, the new data is written to the temporary file and those writes are flushed to disk. Finally, the new file is renamed to the permanent file name, thereby deleting the file with the older data. Under UNIX, rename() is guaranteed to atomically overwrite the old version of the file. Under Windows, the ReplaceFile() call is used to atomically replace one file with another.

In contrast, databases typically offer transactional updates, which allow applications to group many changes into a single atomic unit called a transaction. Complete transactions are applied to the database atomically regardless of system dirty shutdown, database crash, or application crashes. Therefore, applications may safely update their data in whatever manner is most convenient. Depending on the database implementation and the nature of the update, it can be more efficient to update a small portion of the blob instead of overwriting it completely.

The database guarantees transactional semantics by logging both metadata and data changes throughout a transaction. Once the appropriate log records have been flushed to disk, the associated changes to the database are guaranteed to complete. The log is written sequentially, and subsequent database updates can be reordered to minimize seeks. Log entries for each change must be written; but the actual database writes can be coalesced – only the last write to each page need actually occur.

Logging all data and metadata guarantees correctness, but at the cost of writing all data twice. With large data objects, this can put the database at a noticeable performance disadvantage. The sequential write to the database log is roughly equivalent to the sequential write to a file. The additional write to the database pages will form an additional sequential write if all database pages are contiguous or many seek-intensive writes if the pages are not located near each other. Large objects also fill the log causing the need for more frequent log truncation and checkpoint, which reduces opportunities to reorder and combine page modifications.

### 2.3. Data centric web services

To provide an equitable comparison between the filesystem and the database, we burdened the application with detecting and repairing corrupt or torn objects by using SQL Server's bulk-logging mode. Bulk-logging mode provides transactional durability for data in tables, but, like a filesystem, does not guarantee that large object data will be consistent after a dirty shutdown.

Our database and file system based storage systems both provide transactional updates to object metadata. Both write the large object data once and only once. Neither guarantees that large object data are updated correctly across system or application outages.

While handling corrupt objects sounds like a significant burden, the necessary detection and repair mechanisms are often built into large scale web services which use replication. Such services are typically supported by very large, heterogeneous networks built from low cost, unreliable components. System crashes and hardware faults that silently corrupt data are not uncommon. Applications deployed on such systems must regularly check for and recover from corrupted application data. When corruption is detected, intact versions of the affected data can be copied from another replica.

After a crash, a replica could manually check any objects that may have been partially updated when the crash occurred, and obtain good copies from other replicas when necessary. This study does not consider the overhead that recovery mechanisms and detection of silent data corruption would entail. This allows us to provide a meaningful comparison of the performance of the two systems without tying the results to a particular recovery or data scrubbing implementation.

## 3. Prior work

While filesystem fragmentation has been studied within numerous contexts in the past, we were surprised to find relatively few systematic investigations. There is common folklore passed by word of mouth between application and system designers. A number of performance benchmarks are impacted by fragmentation, but these do not measure fragmentation per se. There are algorithms used for space allocation by database and filesystem designers.

We found very little hard data on the actual performance of fragmented storage. Moreover, recent changes in application workloads, hardware models, and the increasing popularity of database systems for the storage of large objects present workloads not covered by existing studies.

### 3.1. Folklore

There is a wealth of anecdotal experience with applications that use large objects. The prevailing wisdom is that databases are better for small objects while filesystems are better for large objects. The boundary between small and large is usually a bit fuzzy. The usual reasoning is:

- Database queries are faster than file opens. The overhead of opening a file handle dominates performance when dealing with small objects.
- Reading or writing large files is faster than accessing large database blobs. Filesystems are optimized for streaming large objects.
- Database client interfaces aren't good with large objects. Remote database client interfaces such as MDAC have historically been optimized for short low latency requests returning small amounts of data.
- File opens are CPU expensive, but can be easily amortized over cost of streaming large objects.

None of the above points address the question of application complexity. Applications that store large objects in the filesystem encounter the question of how to keep the database object metadata and the filesystem object data synchronized. A common problem is the garbage collection of files that have been "deleted" in the database but not the filesystem.

Also missing are operational issues such as replication, backup, disaster recovery, and fragmentation.

### 3.2. Standard Benchmarks

While many filesystem benchmarking tools exist, most consider the performance of clean filesystems, and do not evaluate long-term performance as the storage system ages and fragments. Using simple clean initial conditions eliminates potential variation in results caused by different initial conditions and reduces the need for settling time to allow the system to reach equilibrium.

Several long-term filesystem performance studies have been performed based upon two general approaches [Seltzer]. The first approach, *trace based load generation*, uses data gathered from production systems over a long period. The second approach, and the one we adopt for our study, is *vector based* load generation that models application behavior as a list of primitives (the 'vector'), and randomly applies each primitive to the filesystem with the frequency a real application would apply the primitive to the system.

NetBench [NetBench] is the most common Windows file server benchmark. It measures the performance of a file server accessed by multiple clients. NetBench generates workloads typical of office applications.

SPC-2 benchmarks storage system applications that read and write large files in place, execute large read-only database queries, or provide read-only on-demand access to video files [SPC].

The Transaction Processing Performance Council [TPC] have defined several benchmark suites to characterize online transaction processing workloads and also decision support workloads. However, these benchmarks do not capture the task of managing large objects or multimedia databases.

None of these benchmarks consider file fragmentation.

### 3.3. Data layout mechanisms

Different systems take surprisingly different approaches to the fragmentation problem.

The creators of FFS observed that for typical workloads of the time, fragmentation avoiding allocation algorithms kept fragmentation under control as long as volumes were kept under 90% full [Smith]. UNIX variants still reserve a certain amount of free space on the drive, both for disaster recovery and in order to prevent excess fragmentation.

NTFS disk occupancy on deployed Windows systems varies widely. System administrators' target disk occupancy may be as low as 60% or over 90% [NTFS]. On NT 4.0, the in-box defragmentation utility was known to have difficulties running when the occupancy was greater than 75%. This limitation was addressed in subsequent releases. By Windows 2003 SP1, the utility included support for defragmentation of system files and attempted partial file defragmentation when full defragmentation is not possible.

LFS [Rosenblum], a log based filesystem, optimizes for write performance by organizing data on disk according to the chronological order of the write requests. This allows it to service write requests sequentially, but causes severe fragmentation when files are updated randomly. A cleaner that simultaneously defragments the disk and reclaims deleted file space can partially address this problem.

Network Appliance's WAFL ("Write Anywhere File Layout") [Hitz] is able to switch between conventional and write-optimized file layouts depending on workload conditions. WAFL also leverages NVRAM caching for efficiency and provides access to snapshots of older versions of the filesystem contents. Rather than a direct copy-on-write of the data, WAFL metadata remaps the file blocks. A defragmentation utility is supported, but is said not to be needed until disk occupancy exceeds 90+%.

GFS [Ghemawat], a filesystem designed to deal with multi-gigabyte files on 100+ terabyte volumes, partially addresses the data layout problem by using 64MB blocks called 'chunks'. GFS also provides a safe *record append* operation that allows many clients to simultaneously append to the same file, reducing the number of files (and opportunities for fragmentation) exposed to the underlying filesystem. GFS records may not span chunks, which can result in internal fragmentation. If the application attempts to append a record that will not fit into the end of the current chunk, that chunk is zero padded, and the new record is allocated at the beginning of a new chunk. Records are constrained to be less than ¼ the chunk size to prevent excessive internal fragmentation. However, GFS does not explicitly attempt to address fragmentation introduced by the underlying file system, or to reduce internal fragmentation after records are allocated.

## 4. Comparing Files and BLOBs

This study is primarily concerned with the deployment and performance of data-intensive web services. Therefore, we opted for a simple vector based workload typical of existing web applications, such as Hotmail, flickr, and MSN spaces. These sites allow sharing of objects that range from small text mail messages (100s of bytes) to photographs (100s of KB to a few MB) to video (100s of MBs.)

The workload also corresponds to collaboration applications such as SharePoint Team Services. These applications enable rich document sharing semantics, rather than simple file shares. Examples of the extra semantics include document versioning, content indexing and search, and rich role-based authorization.

Many of these sites and applications use a database to hold the application specific metadata including that which describes the large objects. Large objects may be stored in the database, in the filesystem, or distributed between the database and the filesystem. We consider only the first two options here. We also did not include any shredding or chunking of the objects.

### 4.1. Test System Configuration

All the tests were performed on the system described in Table 1: All test code was written using C# in Visual Studio 2005 Beta 2. All binaries used to generate tests were compiled to x86 code with debugging disabled.

| Table 1: Configurations of the test systems |
|---|
| Tyan S2882 K8S Motherboard, |
| 1.8 Ghz Opteron 244, 2 GB RAM (ECC) |
| SuperMicro "Marvell" MV8 SATA controller |
| 4 Seagate 400GB ST3400832AS 7200 rpm SATA |
| Windows Server 2003 R2 Beta (32 bit mode). |
| SQL Server 2005 Beta 2 (32 bit mode). |

### 4.2. File based storage

For the filesystem based storage tests, we stored metadata such as object names and replica locations in SQL server tables. Each application object was stored in its own file. The files were placed in a single directory on an otherwise empty NTFS volume. SQL was given a dedicated log and data drive, and the NTFS volume was accessed via an SMB share.

We considered a purely file-based approach, but such a system would need to implement complex recovery routines, would lack support for consistent metadata, and would not provide functionality comparable to the systems mentioned above.

This partitioning of tasks between a database and filesystem is fairly flexible, and allows a number of replication and load balancing schemes. For example, a single clustered SQL server could be associated with several file servers. Alternately, the SQL server could be co-located with the associated files and then the combination clustered. The database isolates the client from changes in the architecture – changing the pointer in the database changes the path returned to the client.

We chose to measure the configuration with the database co-located with the associated files. This single machine configuration kept our experiments simple. We avoided building assumptions and dependencies on the network layout into the study. However, we structured all code to use the same interfaces and services as a networked configuration.

### 4.3. Database storage

The database storage tests were designed to be as similar to the filesystem tests as possible. As explained previously, we used bulk-logging mode. We also used out-of-row storage for the application data so that the blobs did not decluster the metadata.

Although the blob data and table information are stored in the same file group, out-of-row storage places blob data on pages that are distinct from the pages that store the other table fields. This allows the table data to be kept in cache even if the blob data does not fit in main memory. Analogous table schemas and indices were used and only minimal changes were made to the software that performed the tests.

### 4.4. Performance tuning

We set out to fairly evaluate the out-of-the-box performance of the two storage systems. Therefore, we did no performance tuning except in cases where the default settings introduced gross discrepancies in the functionality that the two systems provided.

We found that NTFS's file allocation routines behave differently when the number of bytes per file write (append) is varied. We did not preset the file size; as such NTFS attempts to allocate space as needed. When NTFS detects large, sequential appends to the end of a file its allocation routines aggressively attempt to allocate contiguous space. Therefore, as the disk gets full, smaller writes are likely to lead to more fragmentation. While NTFS supports setting the valid data length of a file, this operation incurs the write overhead of zero filling the file, so it is not of interest here [NTFS]. Because the behavior of the allocation routines depends on the size of the write requests, we use a 64K write buffer for all database and file system runs. The files were accessed (read and written) sequentially. We made no attempt to pass hints regarding final object sizes to NTFS or SQL Server.

### 4.5. Workload generation

Real world workloads have many properties that are difficult to model without application specific information such as object size distributions, workload characteristics, or application traces.

However, the applications of interest are extremely simple from a storage replica's point of view. Over time, a series of object allocation, deletion, and safe-write updates are processed with interleaved random read requests.

For simplicity, we assumed that all objects are equally likely to be written and/or read. We also assumed that there is no correlation between objects. This lets us measure the performance of write-only and read-only workloads.

We measured constant size objects rather than objects with more complicated size distributions. We expected that size distribution would be an important factor in our experiments. As shown later, we found that size distribution had no obvious effect on the behavior. Given that, we chose the very simple constant size distribution.

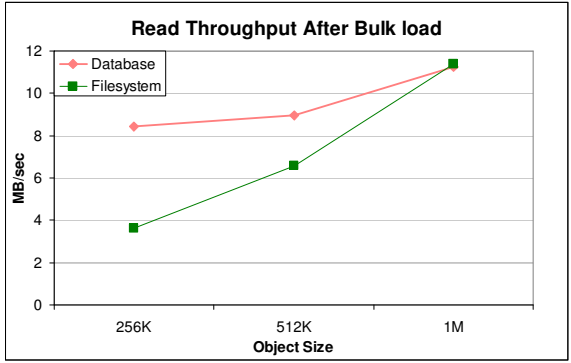

**Figure 1: Read throughput immediately after bulk loading the data. Databases are fastest on small objects. As object size increases, NTFS throughput improves faster than SQL Server throughput.**

### *4.6. Storage age*

We want to characterize the behavior of storage systems over time. Past fragmentation studies measure age in elapsed time such as 'days' or 'months.' We propose a new metric, storage age, which is the ratio of bytes in objects that once existed on a volume to the number of bytes in use on the volume. This definition of storage age assumes that the amount of free space on a volume is relatively constant over time.

For a safe-write system that is the ratio of object insert-update-delete bytes over the number of total object bytes. When evaluating a single node system using trace-based data, storage age has a simple intuitive interpretation.

A synthetic workload effectively speeds up application elapsed time. Our workload is disk arm limited; we did not include extra think time or processor overheads. The speed up is application specific and depends on the actual application read:write ratio and heat.

We considered reporting time in "hours under load". Doing so has the undesirable property of rewarding slow storage systems by allowing them to perform less work during the test.

In our experimental setup, storage age is equivalent to "safe writes per object." This metric is independent of the actual applied read:write load or the number of requests over time. Storage ages can be compared across hardware configurations and applications. Finally, it is easy to convert storage age to elapsed wall clock time after the rate of data churn, or overwrite rate, of a specific system is determined.

## 5. Results

Our results use throughput as the primary indicator of performance. We started with the typical out-of-the-box throughput study. We then looked at the longer term changes caused by fragmentation with a focus on 256K to 1M object sizes where filesystem and database have comparable performance. Lastly, we discuss the effects of volume size and object size on our measurements.

### *5.1. Database or Filesystem: Throughput out-of-the-box*

We begin by establishing when a database is clearly the right answer and when the filesystem is clearly the right answer.

Following the lead of existing benchmarks, we evaluated the read performance of the two systems on a clean data store. Figure 1 demonstrates the truth of the folklore: objects up to about 1MB are best stored as database blobs. Performance of SQL reads for objects of 256KB was 2x better than NTFS; but the systems had parity at 1MB objects, and beyond that NTFS was the best choice.

The write throughput of SQL Server exceeded that of NTFS during bulk load. With 512KB objects, database write throughput was 17.7MB/s, while the filesystem only achieved 10.1MB/s.

### *5.2 Database or Filesystem over time*

Next, we evaluated the performance of the two systems on large objects over time. If fragmentation is important, we expect to see noticeable performance degradation.

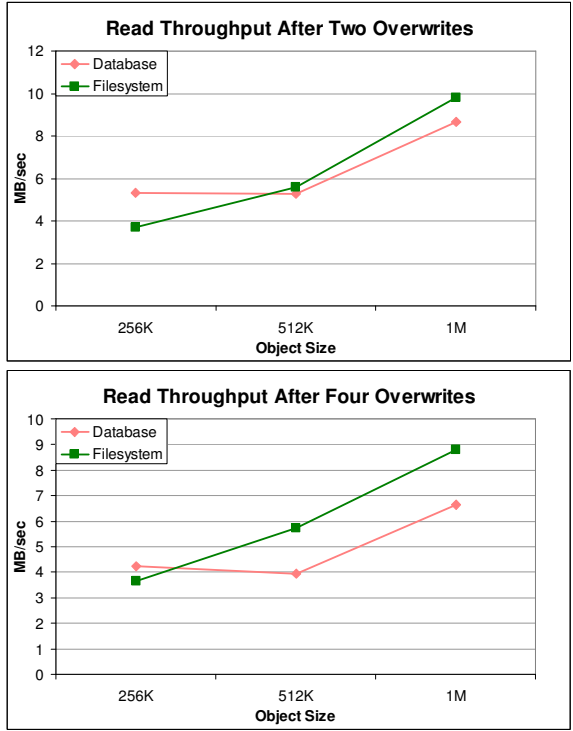

**Figure 2: Fragmentation causes read performance to degrade over time. File system is less affected by this than SQL Server. Over time NTFS outperforms SQL Server when objects are larger than 256KB.**

Our first discovery was that SQL Server does not provide facilities to report fragmentation of large object data or to defragment such data. There are measurements and mechanisms for index defragmentation. The recommended way to defragment a large blob table is to create a new table in large new file group, copy the old records to the new table and drop the old table [SQL].

To measure fragmentation, we tagged each of our objects with a unique identifier and a sequence number at 1KB intervals. We also implemented a utility that looks for the locations of these markers on a raw device in a way that was robust to page headers, and other artifacts of the storage system. In other words, the utility measures fragmentation in the same way regardless of whether the objects are stored in the filesystem or the database. We ran our utility against an NTFS volume to check that it reported figures that agreed with the NTFS fragmentation report.

The degradation in read performance for 256K, 512K, and 1MB blobs is shown in Figure 2. Each storage age (2 and 4) corresponds to the time necessary for the number of updates, inserts, or deletes to be N times the number of objects in our store since the bulk load (storage age 0) shown in

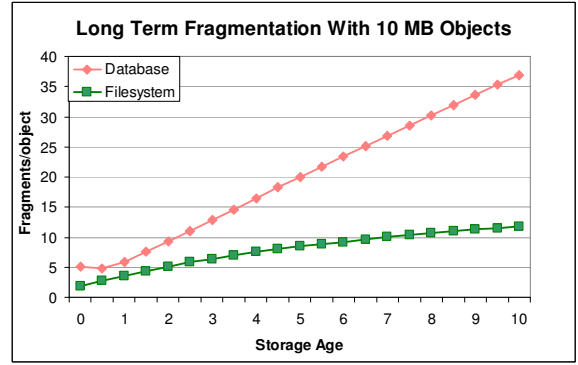

**Figure 3: For large objects, NTFS deals with fragmentation more effectively than SQL server.**

Figure 1. Fragmentation under NTFS begins to level off over time. SQL Server's fragmentation increases almost linearly over time and does not seem to be approaching any asymptote. This is dramatically the case with very large (10MB) objects as seen in Figure 3.

We also ran our load generator against an artificially and pathologically fragmented NTFS volume. We found that fragmentation slowly decreased over time. The best-effort attempt to allocate contiguous space actually defragments such volumes. That experiment also suggests that NTFS is indeed approaching an asymptote in Figure 3.

The degradation in write performance is shown in Figure 4. In both systems, the write throughput during bulk load is much better than read throughput immediately afterward. This is not surprising, as the storage systems can simply append each new file to the end of allocated storage, avoiding seeks during bulk load. On the other hand, the read requests are randomized, and must incur the overhead of at least one seek. After bulk load, the write performance of SQL Server degrades quickly, while the NTFS write performance numbers are slightly better than its read performance.

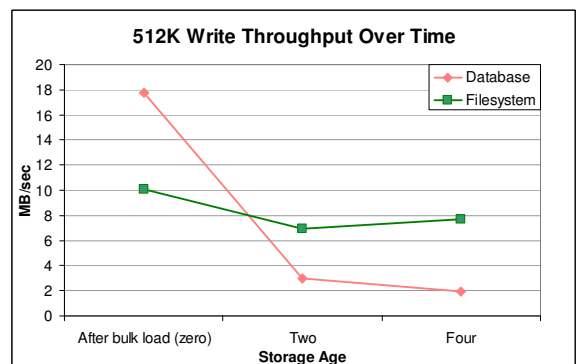

**Figure 4: Although SQL Server quickly fills a volume with data, performance suffers when existing objects are replaced.**

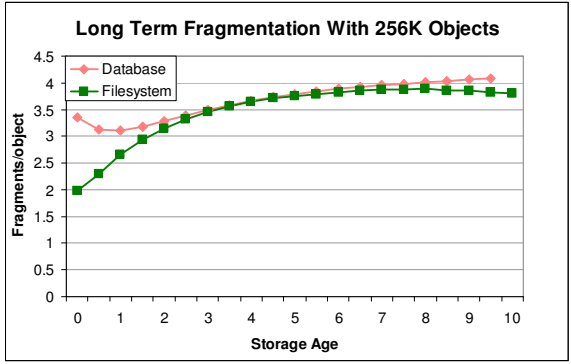

**Figure 5: For small objects, the systems have similar fragmentation behavior.**

Note that these write performance numbers are not directly comparable to the read performance numbers in Figures 1 and 2. Read performance is measured after fragmentation, while write performance is the average performance during fragmentation. To be clear, the "storage age four" write performance is the average write throughput between the read measurements labeled "bulk load" and "storage age two." Similarly, the reported write performance for storage age four reflects average write performance between storage ages two and four.

The results so far indicate that as storage age increases, 256KB, not 1MB is the cross over point where filesystems out perform databases. Objects up to about 256KB objects are best kept in the database; larger objects should be in the filesystem.

To verify this, we attempted to run both systems until the performance reached steady state.

Figure 5 indicates that fragmentation converges to four fragments per file, or one fragment per 64KB, in both the filesystem and database. This is interesting because our tests use 64KB write requests, again suggesting that the impact of write block size upon fragmentation warrants further study. From this data, we conclude that SQL Server indeed outperforms NTFS on objects under 256KB, as indicated by Figures 2 and 4.

### 5.3. Fragmentation effects of object size, volume capacity, and write request size

Distributions of object size vary greatly from application to application. Similarly, applications are deployed on storage volumes of widely varying size particularly as disk capacity continues to increase dramatically.

This series of tests generated objects using a constant size distribution and compared performance when the sizes were uniformly distributed. Both sets of objects had a mean size of 10MB.

Intuition suggested that constant size objects should not lead to fragmentation. Deleting an initially contiguous object leaves a region of contiguous free space exactly the right size for any new object. As shown in Figure 6, our intuition was wrong.

As long as the average object size is held constant there is little difference between uniformly distributed and constant sized objects. This suggests that experiments that use extremely simple size distributions can be representative of many different workloads. This contradicts the approach taken by prior storage benchmarks that make use of complex, accurate modeling of application workloads. This may well be due to the simple all-or-nothing access pattern that avoids object extension and truncation, and our assumption that application code has not been carefully tuned to match the underlying storage system.

The time it takes to run the experiments is proportional to the volume's capacity. When the entire disk capacity (400GB) is used, some experiments take a week to complete. Using a smaller (although perhaps unrealistic) volume size, allows more experiments; but how trustworthy are the results?

As shown in Figure 7, we found that volume size does not really affect performance for larger volume sizes. However, on smaller volumes, we found that as the *ratio of free space to object size* decreases, performance degrades.

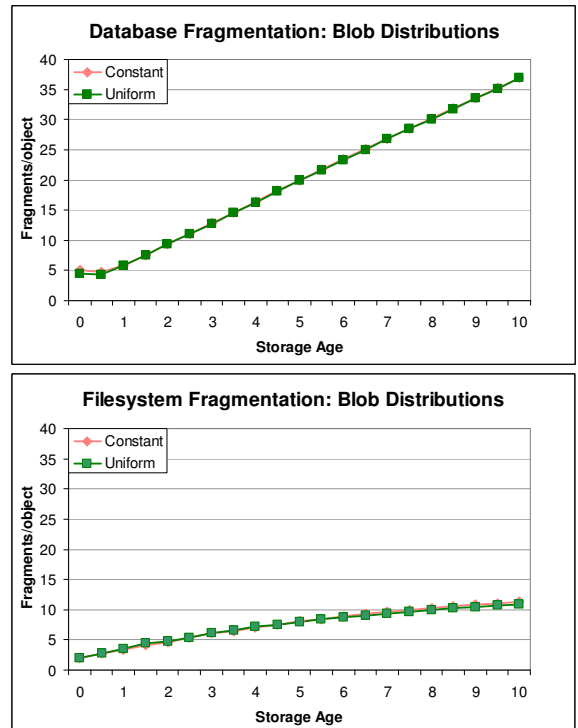

**Figure 6: Fragmentation for large (10MB) blobs – increases slowly for NTFS but rapidly for SQL. However objects of a constant size show no better fragmentation performance than objects of sizes chosen uniformly at random with the same average size.**

We did not characterize the exact point where this becomes a significant issue. However, our results suggest that the effect is negligible when there is 10% free space on a 40GB volume storing 10MB objects, which implies a pool of 400 free objects. With a 4GB volume with a pool of 40 free objects, performance degraded rapidly.

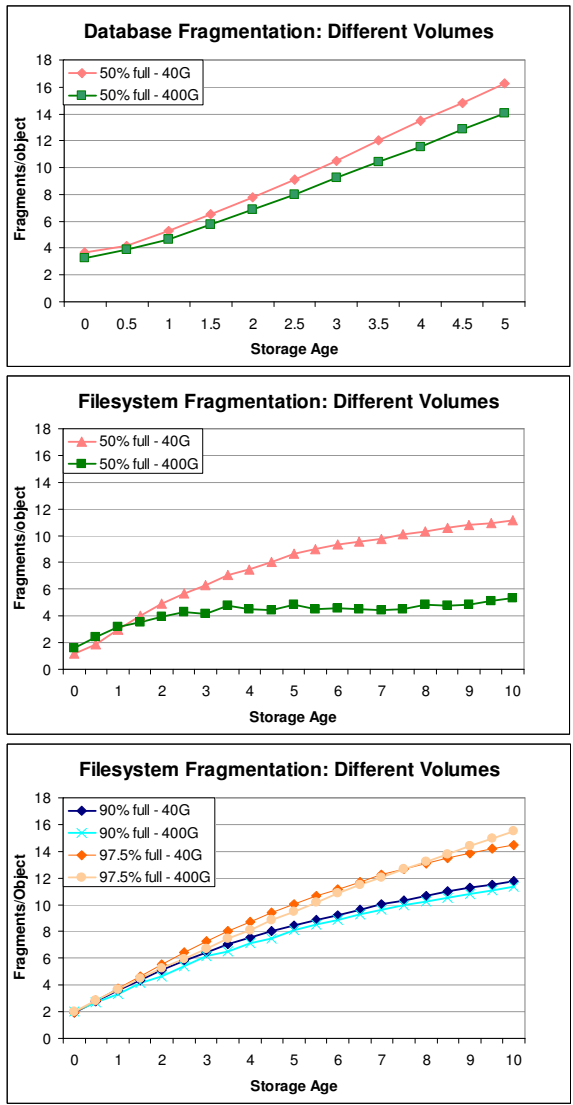

**Figure 7: Fragmentation for 40GB and 400GB volumes. Other than the 50% full file system run, volume size has a negligible impact on fragmentation.**

# 6. Implications for system designers

This article has already mentioned several issues that should be considered during application design. Designers should provision at least 10% excess storage capacity to allow each volume to maintain free space for many (~400 in our experiment) free objects. If the volume is large enough, the percentage free space becomes a limiting factor. For NTFS, we can see this in Figure 7, where the performance of a 97.5% full 400GB volume is worse than the performance of a 90% full 40GB volume. (A 99% full 400GB volume would have the same number of free objects as the 40GB volume.)

While we did not carefully characterize the impact of application allocation routines upon the allocation strategy used by the underlying storage system, we did observe significant differences in behavior as we varied the write buffer size. Experimentation with different buffer sizes, or other techniques that avoid incremental allocation of storage may significantly improve long run storage performance. This also suggests that file system designers should re-evaluate what is a "large" request and be more aggressive about coalescing larger sequential requests.

Simple procedures such as manipulating write size, increasing the amount of free space, and performing periodic defragmentation can improve the performance of a system. When dealing with an existing system, tuning these parameters may be preferable to switching from database to file system storage, or vice versa.

When designing a new system, it is important to consider the behavior of a system *over time* instead of looking only the performance of a clean system. If fragmentation is a significant concern, the system must be defragmented regularly. Defragmentation of a filesystem implies significant read/write impacts or application logic to garbage collect and reinstantiate a volume. Defragmentation of a database requires explicit application logic to copy existing blobs into a new table. To avoid causing still more fragmentation, that logic must be run only when ample free space is available. A good database defragmentation utility (or at least good automation of the above logic including space estimation required) would clearly help system administrators.

Using storage age to measure time aids in the comparison of different designs. In this study we use "safe-writes per object" as a measurement of storage age. In other applications, appends per object or some combination of create/append/deletes may be more appropriate.

For the synthetic workload presented above, filesystem based storage works well for objects larger than 256KB. A better database blob implementation would change this. At a minimum, the database should report fragmentation. An in-place defragmentation utility would be helpful. To support incremental object modification rather than the full rewrite considered here, a more flexible B-Tree based blob storage algorithm that optimizes insertion and deletion of arbitrary data ranges within objects would be advantageous.

## 7. Conclusions

The results presented here predict the performance of a class of storage workloads, and reveal a number of previously unknown factors in the importance of storage fragmentation. They describe a simple methodology that can measure the performance of other applications that perform a limited number of storage create, read, update, write, and delete operations.

The study indicates that if objects are larger than one megabyte on average, the file system has a clear advantage. If the objects are under 256 kilobytes, the database has a clear advantage. Inside this range, it depends on how write intensive the workload is, and the storage age of a typical replica in the system.

Instead of providing measurements in wall clock time, we use storage age, which makes it easy to apply results from synthetic workloads to real deployments.

We are amazed that so little information regarding the performance of fragmented storage was available. Future studies should explore how fragmentation changes under load. We did not investigate the behavior of NTFS or SQL Server when multiple writes to multiple objects are interleaved. This may happen if objects are slowly appended to over long periods of time or in multithreaded systems that simultaneously create many objects. We expect that fragmentation gets worse due to the competition, but how much worse?

## 8. Acknowledgements

We thank Eric Brewer for the idea behind our fragmentation analysis tool, helping us write this article and reviewing several earlier drafts. We also thank Surendra Verma, Michael Zwilling and the SQL Server and NTFS development teams for answering numerous questions throughout the study.